# Exploring Risk and Fraud Scenarios in Affiliate Marketing Technologies from the Advertiser's perspective

**Bede Ravindra Amarasekara**
School of Engineering and Advanced Technology
Massey University
Albany, New Zealand
Email: bede@amarasekara.net

**Anuradha Mathrani**
School of Engineering and Advanced Technology
Massey University
Albany, New Zealand
Email: a.s.mathrani@massey.ac.nz

## Abstract

Affiliate Marketing (AM) has become an important and cost effective tool for e-commerce. There are numerous risks and vulnerabilities that are typically associated with AM. Though a well-planned AM model can greatly benefit the e-commerce strategies of an enterprise, a haphazardly implemented system can expose a business enterprise to major risks and vulnerabilities, which can lead to great financial losses through fraudulent activities. This research has identified some of the risks and the technical background of those scenarios. In-depth investigation of a bespoke application used for reconciling transactions between an AM network and the back-office systems of an advertiser uncovered many risks and vulnerabilities. Next, a functional prototype of an AM network comprising of an e-commerce website of an advertiser, an affiliate website and an AM platform has been developed. This paper discusses the simulations of fraud scenarios and risky settings via the functional prototype.

## 1 INTRODUCTION

Businesses such as those in the Travel and Tourism sector, having customers at far away locations, are heavily dependent on e-commerce for revenue generation (Mariussen, Daniele, & Bowie, 2010; Gregori, Daniele, & Altinay, 2013). These companies register on search engines (e.g., Google, Bing, Yahoo), with the aim of gaining higher visibility through rankings defined by search engine optimization algorithms. Apart from search engine visibility, online advertising using tools such as Google Ad-words in conjunction with Google Analytics and Universal Analytics are important on-line marketing strategies.

"Affiliate marketing" (AM) is fast becoming another main strategy in on-line marketing (PR Newswire, 2015). CDNow is considered the early pioneer in AM, starting in 1994 (Hoffman & Novak, 2000; Fiore & Collins, 2001), while Amazon.com has the most successful affiliate program with over one million members worldwide (Fox & Wareham, 2012). Venugopal, Das and Nagaraju (2013) traces AM's origin further back in time, to William J. Tobin, the founder of PC Flowers and Gifts launched on the Prodigy Network in 1989, generating more than 6 million dollars by 1993. Dennis and Duffy (2005) recognized a decade ago, the potential of online marketing through AM strategies. Since then the popularity of AM has grown tremendously over the years (Fox & Wareham, 2012; Gregori, Daniele, & Altinay, 2013). Instead of the earlier trend of finding products that aligned well with the topic of the website, now affiliates first look for best-selling products offered by advertisers through affiliate channels, and then build new websites, choosing a topic that would attract potential customers to the site (Collins, 2011a). Figure 1 shows the results of a survey carried out by Collins (2011b) among over one thousand four hundred affiliates with the question, how they usually find out about an Affiliate program to join.

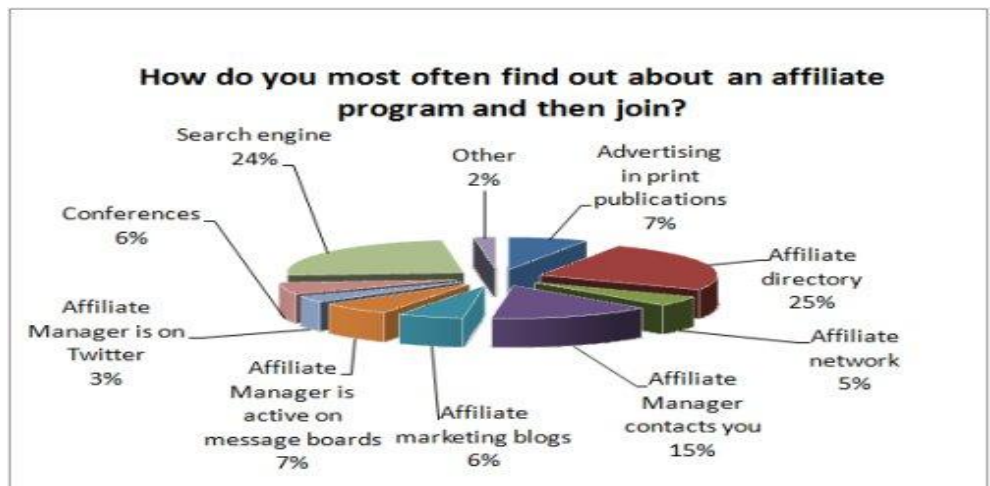

*Figure 1: How to find an Affiliate program to join? (Source: Collins, S. 2011b)*

However, online environments lack visibility and questions are raised over unknown features that may be embedded within the web infrastructure, causing much concern to both businesses and consumers. The perception of risks during transactions is high for both business-to-business and business-to-consumer transactions. Usually in AM programs, affiliates and businesses have never seen each other and have a very limited knowledge of each other's businesses and reputations. Trust is an important factor among all the stake holders associated with the AM value chain. There has been much academic interest on AM from different perspectives, such as the consumer's perspective (Gregori, Daniele and Altinay, 2013), affiliates perspective (Benedictova and Nevosad, 2008) and the merchant's or businesses' perspective (Fox and Wareham, 2012; Libai, Biyalogorski and Gerstner, 2003; Gregori, Daniele and Altinay, 2013). McKnight, Choudhury and Kacmar (2002) identify trust as the key factor between the consumer and the websites.

This research is undertaken to identify potential risks and vulnerabilities associated in AM from a merchant's perspective. The paper first discusses how affiliate marketing platforms operate and gives a brief overview of the technological platform. Next, we explore vulnerabilities which could lead to possibilities of fraud from the merchant's perspective. A functional prototype of an AM network comprising of an e-Commerce website of an advertiser, an affiliate website and an AM platform has been developed. The paper proceeds to expose some of the risks uncovered during our investigation, which will be addressed in future stages of this study.

## 2　TECHNICAL OVERVIEW OF AN AFFILIATE MARKETING PLATFORM

An Affiliate Marketing platform connects an e-commerce site of a merchant (called an advertiser in AM), with a network of affiliates. Affiliates are third-party, independent websites, such as travel blogs, information pages on diverse subjects and interest groups, etc. who are willing to display advertisement links of different advertisers for a monetary gain. The affiliates will drive consumer traffic to an advertiser's e-commerce site, with the expectation that some of those site visitors might convert visits to sales and become customers of the advertiser. In return, an affiliate earns a commission from the advertiser. Though CPC (Cost per Click) advertising model was common at the beginning of e-commerce era, currently CPA (Cost per Acquisition) model is more favoured by some advertisers, because an affiliate is paid a commission, only if the visitor actually makes a purchase. CPC is an advertising model where affiliates are paid a set fee, for traffic generated by each visitor's click. Under CPA affiliates are paid a fee for only visits that converted to an income bearing activity such as a booking or a purchase. A glossary of terminology used by AM and Performance Marketing industry, is compiled by Cake Performance Marketing (Performance Marketing Glossary, 2015). The AM terminology can be confusing and is not always commonly or widely understood, as they can have different meanings under various contexts.

Usually, managing affiliates and tracking transactions while minimising fraud is a specialised activity which is outsourced to a third-party Affiliate Marketing/Performance Marketing Management Platform. Such specialised platforms have technical capabilities to track transactions across the affiliate network and reporting capabilities that keep affiliates and advertisers up to date on their performance and amounts of revenue and commissions earned. Therefore, advertisers, affiliates, customers and the Affiliate Management Platform (AMP) are four parties that are stakeholders of AM.

Due to the lack-of-trust between advertisers and affiliates, performance monitoring is an important and integral part of AM. Advertisers face risks of possible fraudulent activities by affiliates claiming payments for visitor traffic that they did not generate or claiming commissions on sales, for site visits that eventually did not convert to sales. Equally, affiliates are wary of advertisers who might wilfully omit some of the sales transactions that should have earned a commission for the affiliate, by deliberately not recording some tracking information, or adjusting sales volumes, etc.

Affiliates can measure the volume of traffic their websites have generated towards each advertiser's e-commerce site, by monitoring and logging each click-action by visitors to affiliate's websites. But, if any of those clicks converted to a sale and if so, the value of such sale would not be known to the affiliate, unless the advertiser is willing to implement tracking code, that would notify an affiliate in real-time, of these transactions, as and when they happen. Though the advertiser's tracking code will notify the sales data to the AMP in real-time, it is not the usual case, to notify an affiliates system. Hence, the affiliate relies upon the trust, hoping that the advertiser would be honest in declaration of the transactions.

An advertiser can either manage their affiliate program themselves or outsource this activity by subscribing to an AMP.

## 2.1 Affiliate Management and Tracking by the Advertiser

Advertisers can maintain logs of the traffic generated by individual affiliates and also logs of sale values that were generated by affiliate driven traffic. The advertiser can make this information available to the affiliates, either through personalised web pages dedicated to each affiliate, or by email transaction statements sent periodically (daily, weekly or monthly, etc.). These can be as elaborate as an affiliate portal with dedicated personalised pages consisting of comprehensive reports and charts or be simple periodical transaction statements sent by email. An elaborate portal would demand a fair amount of web site development and maintenance for the advertiser.

The advantage would be saving the costs of subscribing to a third-party AMP and information security. Ray (2001) states that the perceived advantage of a lower affiliate management cost by running the process in-house, needs a large enough volume of sales. On the other hand, the affiliate has to trust the honesty of the advertiser, as there is no way for the affiliate to verify the data. The lack of trust between affiliates and advertisers causes this to be a disadvantage.

## 2.2 Affiliate Management and Tracking by an Affiliate Management Platform

The service of specialised third-party platforms are readily available to advertisers, based on a monthly subscription. Outsourcing to these platforms alleviates the need for advertisers to develop and maintain additional functionality to track affiliate activities and the need to design elaborate reporting facilities within their systems. Subscription based AMPs are developed and maintained by external service providers with the latest technologies. Further, trust is one of the core advantages from the affiliate's point of view. Ray (2001) finds the ability of such a platform to introduce new affiliates to an advertiser, with whom they are already working, an added advantage for the advertiser. But, most of all, the ability to provide unbiased transaction and performance data, which works like a third-party audit, helps to win affiliate's trust, that the data has not been manipulated by the advertiser. This gives a greater credibility to the Affiliate Marketing program of the advertiser. The main disadvantages are the subscription cost and the compromising of information security. The third-party AMP receives all the detailed transaction information of not only the online sales that was driven by affiliates, but every online sale of the company, including direct sales through organic searches, search engines and direct URL visits.

## 3 CASE SCENARIO

An investigation of the functionality of an Affiliate Management Network was carried out. An advertiser (a rental car company operating in New Zealand and Australia), a few affiliates connected to this advertiser and a third-party performance marketing/affiliate marketing platform that was used by the advertiser were studied. A prototype has been developed to simulate an Affiliate Network scenario. We have used this prototype to simulate some of the risks and frauds we discovered, in order to further investigate and find solutions to mitigate these risks. The scenario used in this study is described next.

A car rental company "Best Cars" (pseudonym) in New Zealand, has implemented an AM program by signing up with affiliate management Platform provider "Connex" (pseudonym) from USA. Connex tracks all the visitor traffic to Best Cars that are generated by users clicking on links at affiliate sites. Connex also tracks all visitors who make an actual booking at Best Cars. When a booking is made, Connex calculates the commission amount for the affiliate who generated that click. "Globetrotter" (pseudonym) is a popular travel blog based in Hong Kong. It has a New Zealand travel page, which lists many interesting travel tips and latest travel-related offers available in New Zealand. Globetrotter is an ideal web site to become an affiliate for Best Cars, as readers of New Zealand travel pages on Globetrotter might be planning a trip to New Zealand, and might be interested to know of best car rental deals in New Zealand. Therefore, there are three stakeholders (1) an advertiser who wants to sell a product, (2) an affiliate who would use his client base (visitors to affiliates web site) to promote the product, and (3) an AMP that uses the technology to track all the traffic from affiliate to advertiser. A typical affiliate network would have many hundreds or thousands of such affiliates promoting the product of the advertiser.

A "Click-Pixel", which is either an image (banner advertisement) or a text phrase such as "For best deals in car rentals…" within a paragraph of body text, was placed on Globetrotter's web page, with a hyperlink that points to the Connex Platform. The hyperlink contains affiliate id, advertiser id, offer id, and as many data fields an affiliate needs to pass to Connex for tracking, in its request parameters.

On Best Car's payment confirmation page, a "Conversion Pixel" was placed, which is a piece of JavaScript code or in case JavaScript is disabled in the browser, a resource request from Connex web server, with data fields such as session id, total price, etc. as parameters. The session id will help Connex to track the sales conversion to the matching user click accurately.

In addition to the two main tasks of click-tracking and conversion-tracking, the Connex platform can typically provide additional features such as a customised Web Portals for the Best Cars to manage their AM Systems, affiliate commission payment on behalf of Best Cars and sourcing new affiliates and managing them to expand the affiliate network.

## 4   TRACKING PROCESS

There are two parts to the tracking process: click-tracking (to record a visitor's mouse-click-action at Globetrotter website) and conversion-tracking (to record an action, such as a reservation that happened at the Best Car's web site). Figure 2 shows the complete tracking process of a specific transaction, which involves the process of matching a conversion-tracking against a corresponding click-tracking. This process allows the AMP to ascertain which affiliate's site generated that traffic, so that the affiliate can be rewarded with the agreed financial gain, in form of a commission payment. If a conversion-tracking does not have a corresponding click-tracking, such transaction would have occurred through a customer typing the URL of the advertiser directly in to the customer's browser or through an organic search or a paid search, but not through an affiliate's web site.

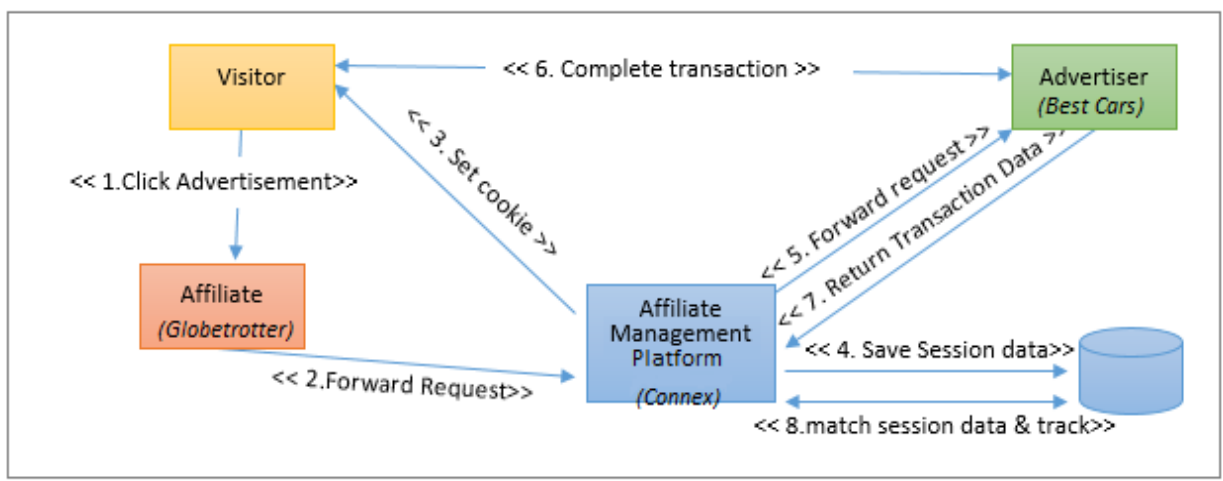

*Figure 2: Tracking process*

## 4.1 Click-tracking

A visitor to the Globetrotter's New Zealand travel blog page clicks on the banner advertisement, which has a URL that points to Connex, instead of pointing to the Best Car's landing page. Connex will place one or more cookies, which are usually known as "tracking cookies", in the visitor's cookies folder of the computer. These cookies contain information that Connex would use for tracking and identification purpose, including a unique Session ID that would identify the visitor again at Best Cars website, on subsequent visits. It will create a Session Cookie with the same Unique ID, which could be used for tracking in case the browser has disabled cookies, which will only last during the current session. After saving information that was passed as URL parameters in to a database, Connex will forward the request to the correct product page (landing page) of the advertiser.

## 4.2 Conversion-tracking

The visitor usually does not notice that the request to the advertiser's product page has been routed via an AMP, because the request forwarding happens transparently to the customer, at a very high speed. When the site visitor makes a booking, during the confirmation process, the Conversion-Pixel placed on the advertiser's confirmation page will pass the sales data such as session id, transaction id and total price, back to Connex. If the site visitor leaves the advertiser's site without making a booking, the "Conversion-Tracking" will not occur and it will be marked by Connex as a "click" (visit), only. If the site visitor returns to Best Cars website and complete a transaction at a later date, the conversion can be only matched against the first visit through session id and the affiliate will earn his commission, only if the cookie that was placed during click-tracking is still valid. Even if the site visitor has disabled cookies, the session cookie that was placed during the click-tracking process, will be still active and available if the site visitor completes the booking transaction within the same session, before closing the web browser.

## 5 FINDINGS: RISKS AND VULNERABILITIES

An in-depth examination of the processes of a bespoke AMP integration & reconciliation application was undertaken. This application reconciles transactions recorded in the AMP with the back-office databases of the advertiser, using the Web API of the AMP. Such an application is an effective first step in minimising fraudulent activities, as the application filters out any fraudulent sales transactions that do not have a corresponding actual transaction record in the back-office databases of the advertiser. Such filtered-out transaction data include bogus sales transactions that never occured, which a rogue affiliate might have caused to enter in to the AMP or falsely altered transaction amounts to earn higher commissions, etc. The data from AMP showed all the clicks (website visits) and the conversions (visits converted to bookings). The conversion data included the total revenue earned by the booking and a booking reference number. As this advertiser implemented a "Revenue-share" advertising model, affiliates were paid a commission only for "conversions". Hence, we were aware that the AMP reported conversions might include genuine conversions as well as possible fraudulent actions that imitated a conversion. If the advertiser implemented a CPC model, we could have seen many more fraudulent activities that would have caused an increase in the amount of clicks (visitor traffic), because the affiliates are paid for each click. It is often easier to manipulate click data than conversion data, hence CPC model is prone to more fraud activity than CPA. Nevertheless, click data has also been examined to uncover any patterns or signs of fraudulent activities. AMP generated web data and the resulting log files were examined, which enabled us to isolate affiliate generated fraudulent web sales data and possible errornous data caused due to bugs or limitations of technology.

It is suggested that around 1% visitors to an affiliate site actually clicks on an advertisement link (Benediktova & Nevosad, 2008). A rogue affiliate can instead cause, what we call a "*Load-time click*" by adding the trigger-code on "page load" event, instead of "click event" thus mimicking, as if every visitor has clicked a link, or clicked all the advertisement-links in the page. Though the visitor might not have clicked on any advertisement, the vistior would not be aware of the 'behind-the-scene" clicks. This fraud is easier to implement on CPC model, which only tracks the clicks. A CPA model needs to use cookies for conversion-tracking purpose, hence, without an "Access-Control-Allow-Origin:[domain name]" header from AMP's web server, CORS (Cross Origin Resource Sharing) restrictions on cross-site scripting (XSS) prevents Load-time clicking, when the browser and server need to exchange cookies. Therefore, in case of CPA scenarios, "*JSONP" (JavaScript Object Notation with Padding)* or an embedded "*iframe*" can be used to achieve a similar result. This fraud can effectively be used, in combination with the "Cookie stuffing" method discussed below. It can insert a cookies from diverse advertisers, in to a user's computer, without any visual clue to the computer user. We made the above findings, while simulating on our AMP prototype.

Edelman and Brandi (2015) have discussed currently known affiliate frauds in considerable detail, some of which are mentioned below. For the purpose of their paper, they have limited the discussion of technology, by stating that browser "cookies" are the fundamental feature that enables tracking. In this paper, we expand the topic further by looking into the technical aspects of cookie tracking and fraud scenarios. We also present further fraud scenarios that we discovered in the process of developing the integration application and during simulation sessions using the prototype.

*Cookie stuffing* discussed by Edelman & Brandi (2015) is a process, in which an affiliate will place many different cookies belonging to different advertisers (third-party cookies), in the visitor's computer. If the visitor subsequently visits any of those advertisers' sites, and makes purchases, the affiliate will earn a commission, without actually having taken part in leading the visitor to that site. The fraud can be combined with "Load time click" fraud, to maximise the commission earnings by an affiliate. The lifespan of the cookie is an imperative factor, which is often an individual business decision of each advertiser.

An affiliate can do, what we call a *"Conversion hijack"* by monitoring the activity of a direct visitor who did not come through affiliate's site, but who is about to make a purchase. By triggering a click-pixel just before the purchase, the affiliate places a cookie on visitor's computer and hijacks the conversion to earn the commission for a sale that would have happened, anyway. Such monitoring can be done using Adware (Edelman & Brandi, 2015) or similar malicious software installed on a visitor's computer, by affiliate, which is an external threat. An internal threat, that could cause such monitoring to happen would be when an affiliate gets the help of a staff member of adveriser's company with sufficient rights, to embed a small code segment on the Web Server, as we discovered in our investigation.

Apart from technology based frauds mentioned above, rogue affiliates can undertake other fraudulent activities that can incur heavy losses using simple deceitful actions. On most e-commerce sites, where a consumer books a service weeks or months ahead of time, such as at tourism related e-commerce sites, an affiliate can book a car or a room himself, a few months ahead and eventually cancel the booking, earning a hefty commission. The AMP integration application that we examined for the purpose of this paper was capable of effectively controlling this category of fraud, by reconciling AMP conversion records with the back-office databases of the advertiser.

From the advertiser's perspective, some threats originate from external sources such as, by affiliates, site visitors and hackers, etc., while others are internal threats attributed to advertiser's staff with appropriate security access levels, contractors, IT service providers, etc. Internal threats can be more severe as internal staff can have unrestricted access to, and a comprehensive knowledge of, the IT systems. The most significant risk we established and subsequently tested using our prototype, is what we call *"conversion stealing"*. It is the process of selecting legitimate transactions from advertiser's back-office databases, which did not originate through any affiliates, e.g. direct traffic or traffic generated through search engines, and creating tracking entries on the AMP servers, attributing them to a sepcific rogue affiliate, who will earn the commission. As many advertisers can have more than half of the on-line sales originating from sources other than affiliate marketing, "conversion stealing" can lead to large losses for an advertiser. This risk can originate internally or externally, though it is very much more easier to implement from within the organisation of the advertiser. Even an AMP integration application might not be able to detect this fraud, unless specifically designed to handle this threat, because these illegitimate conversion trackings, in fact refer to legitimate transactions within the advertiser's system.

### 5.1 Tracking failures

Tracking failures are not fraudulent activities, nevertheless can cause losses to the affiliates losing their rightfully earned commissions. Usually, saving of cookies is allowed by the client's browser, unless otherwise explicitly disabled by the user.

Even if cookies are disabled, due to the use of Session cookies, which have a life span of the current session, the tracking will function correctly, as long as the user does not close the browser, between the click-tracking process and conversion-tracking process.

Our prototype simulations have enabled us to recognise some situations under which tracking will fail:

- If the user clears the browser cache between the click-tracking process and conversion-tracking process.

- If the user uses the "incognito mode" or similar "*private*" browsing feature offered by most browsers, and does not complete the "conversion" in the same browsing session, instead closes the browser, and starts another session of the browser to continue the purchase. But, if the user completes the conversion, then the tracking process will be successful, as session cookie, which has the lifespan of the session duration, will be used.

- If the cookie has expired when the user returns to complete the purchase.

- If the visitor has more than one browser installed on his/her computer, and uses one browser to visit the affiliate's web site and later uses another browser on the same computer to navigate directly to the advertiser's website and makes a purchase. The second browser has no access to the cookie storage of the first browser, as browsers do not share cookies stored, between them, the conversion will not be tracked.

- If the user uses two different computers, one computer to browse the affiliate's website, but uses a different computer to navigate directly to the advertiser's website to make a purchase. Again the conversion will not be tracked, as the cookie was placed in a different computer.

## 5.2　Information security

Though not associated with direct loss of revenue, compromising information security is a significant vulnerability in an AM network. Our analysis of the AMP log files have enabled us to discover numerous threats to information security, due to tracking activity, which we were able to further investigate through simulation using the AMP prototype. Due to lack of understanding of technologies that drive an AM system, most AM practitioners are not fully aware to what extent, the privacy of their critical business data are compromised. When the conversion-tracking pixel is placed on the advertiser's confirmation page, each time a customer completes a transaction, the pixel fires the information back to the AMP, which in turn would match the conversion with a click-tracking. If there is no matching click-tracking found, the commission is not paid to any affiliate, as it would be a either a direct sale that did not generate through an affiliate site, or it could be due to any of the reasons discussed under the "Tracking failure" section. Many advertisers are not aware that the AMP has a running sum of each electronic transaction or booking that has occurred on the advertiser's web site.

If the advertiser chose to use a tagging service such as Google analytics or Universal Analytics, on top of the AMP for pixel tracking, then the advertiser exposes all the electronic business transaction data to Google, as well as to the Affiliate Marketing Platform. This could be in breach of privacy laws in some countries in the world. Hence, Google requests all Universal Analytics users to display a notice to the effect that cookies are used to track specific information about users, which would indemnify Google from privacy infringement accusations.

# 6　DISCUSSION

Literature review reveals that individual perceptions and definitions of fraudulent activities are somewhat subjective. What some companies consider as an illegal activity can be quite legitimate with another company, for instance, keyword and trademark bidding or "Typo squatting" is illegal with some advertisers, but not with others. Edelman & Brandi (2015) have raised some risks associated with AM. As cookies are widely used for tracking, the manner of usage of cookies is a very important business decision, that differs from company to company. An advertiser may consider it to be a USP (Unique Selling Point) to have multiple affiliate references, and share the commission between all affiliates, while another advertiser may decide to offer the commission to the last affiliate who referred the same customer. A cookie with a long life time or one that does not expire till next conversion of sale, would allow an affiliate to earn the commission, even if the customer returns to purchase after a long lapse since the first visit. In case of Amazon, the cookie expires in a few hours, allowing an affiliate only a small window of opportunity to earn their commission. However, if the customer added the item to a shopping cart and later abandoned it, but then at a later date decides to complete the purchase, the affiliate will have a seven-day window (Ray, 2001). Hence, technology can be used to minimise identified risks and vulnerabilities, but the solution should be flexible enough to allow it to be implemented to suit different business decisions depending on different business models. The AMP can offer these fine-grained controls as advanced features within their tracking system, for the advertisers to control.

Advertisers who sell services that are usually booked many months ahead such as travel and tourism related services, need to implement a transaction reconciliation application. These integration applications reconcile AMP's conversion tracking information with advertiser's back-end transaction databases. This can prevent affiliates from first collecting commissions and subsequently cancelling reservations.

# 7  CONCLUSION AND FUTURE DIRECTION

Though a well-planned affiliate marketing model can greatly benefit the e-commerce strategies of an enterprise, a haphazardly implemented system can expose a business enterprise to major risks and vulnerabilities, which can lead to great financial losses through fraudulent activities.

Our prototype will provide a proof of concept to demonstrate the possibilities of fraud within the existing AMP infrastructure. As further fraud scenarios are identified, we will selectively design appropriate technological solutions to address them. We hope with our prototype, we can isolate some technological glitches and implement solutions to reduce chances of risk and fraud in existing affiliate marketing platforms.

# 8  REFERENCES

Benediktova, B., & Nevosad, L. (2008). *Affiliate Marketing - Perspective of content providers*. Department of Business Administration and Social Sciences, Lulea University of Technology.

Collins, S. (2011a). Finding Affiliate Programs. Retrieved July 27, 2015, from http://www.extramoneyanswer.com/finding-affiliate-programs

Collins, S. (2011b). *Affiliate summit 2011*. Retrieved from ISSUU: http://issuu.com/affiliatesummit/docs/affstat2011report

Dennis, L., & Duffy. (2005). Affiliate marketing and its impact on e-commerce. *Journal of Consumer Marketing, 22*(3), 161-163.

Edelman, B., & Brandi, W. (2015, February). Risk, Information, and Incentives in Online Affiliate Marketing. *Journal of Marketing Research, LII*, 1-12.

Fiore, F., & Collins, S. (2001). *Successful Affiliate Marketing for Merchants*. Indianapolis, IN: Que Corp.

Fox, P. B., & Wareham, J. D. (2012). Governance Mechanisms in Internet-Based Affiliate Marketing Programs in Spain. In I. Lee, *Transformations in E-Business Technologies and Commerce: Emerging Impacts* (pp. 222-239). doi:10.4018/978-1-61350-462-8

Gregori, N., Daniele, R., & Altinay, L. (2013, June 18). Affiliate Marketing in Tourism: Determinants of Consumer Trust. *Journal of Travel Research*. doi:10.1177/0047287513491333

Hoffman, D. L., & Novak, T. P. (2000). How to acquire customers on the web. *Harvard business review*, 179-188.

Libai, B., Biyalogorsky, E., & Gerstner, E. (2003, May). Setting Referral Fees in Affiliate Marketing. *Journal of Service Research, 5*(4), 303-315. doi:10.1177/1094670503005004003

Mariussen, A., Daniele, R., & Bowie, D. (2010). Unintended consequences in the evolution of affiliate marketing networks: a complexity approach. *The Services Industries Journal*, 1707-1722. doi:10.1080/02642060903580714

McKnight, H. D., Choudhury, V., & Kacmar, C. (2002, Sep.). Developing and Validating Trust Measures for e-Commerce: An Integrative Typology. *Information Sysstems Research, 13*(3), 334-359.

*Performance Marketing Glossary*. (2015, May 28). Retrieved July 20, 2015, from Cake Performance Marketing: http://getcake.com/performance-marketing-glossary/

PR Newswire. (2015, January 19). *First Performance Marketing Offering to Feature Real-time Fraud Detection Powered by Industry Leaders Cake by Accelerize and Forensiq*. PR Newswire US.

Ray, A. (2001, August 23). Affiliate Schemes prove their worth. *Marketing - London*, pp. 29-30.

Venugopal, K., Das, S., & Nagaraju, M. (2013, June). Business Made Easy By Affiliate Marketing. *Journal of Business Management & Social Sciences Research, 2*(6), 50-56.

## Copyright